# A Goal-oriented Framework for Data Warehousing Quality Measurement

Cristina Cachero and Jesús Pardillo

Department of Software and Computing Systems
University of Alicante, Spain
`{ccachero,jesuspv}@dlsi.ua.es`

**Abstract.** Requirements engineering is known to be a key factor for the success of software projects. Inside this discipline, *goal-oriented requirements engineering* approaches have shown specially suitable to deal with projects where it is necessary to capture the alignment between system requirements and stakeholders' needs, as is the case of data-warehousing projects. However, the mere alignment of data-warehouse system requirements with business goals is not enough to assure better data-warehousing products; measures and techniques are also needed to assure the data-warehouse quality. In this paper, we provide a modelling framework for *data-warehouse quality measurement* ($i^*$DWQM). This framework, conceived as an $i^*$ extension, provides support for the definition of data-warehouse requirements analysis models that include quantifiable quality scenarios, defined in terms of well-formed measures. This extension has been defined by means of a UML profiling architecture. The resulting framework has been implemented in the *Eclipse* development platform.

## 1 Introduction

Data-warehouse systems provide a multidimensional view of heterogeneous operational data sources in order to supply valuable information to decision makers. Its development is usually based on the multidimensional modelling, because of its intuitiveness and its support for high-performant queries [9, 10]. Since the data-warehouse integrates several operational data-sources, the design of multidimensional models has been traditionally guided by supply-driven approaches [7, 8]. However, in order to assure the adequation of such designs to the information needs of decision makers, a requirement-analysis stage is needed. For this stage, goal-oriented frameworks have proven specially suitable. The reason for this fact is twofold: First, goal-oriented frameworks provide constructs for the modelling of large organisational contexts, which are the commonality in data-warehouses. Second, they match the way in which decision makers express themselves, *i.e.*, in terms of general expectations or objectives that the data-warehouse should support. This suitability has been materialised in proposals such as $i^*$DWRA [12] or the one presented in [6].

However, the inclusion of goals, although necessary, may not be sufficient to guarantee the quality of data-warehouse systems. Indeed, although a good methodology with accurate goal definitions may lead to good and suitable data-warehouse models, many other factors could influence their quality, such as human decisions. It is thus necessary to complete data-warehousing methodologies with measures and techniques for product quality assessment [16, 17, 2, 15]. One of the best known techniques in this sense, emphasised in well-known software development processes such as the *unified process* (UP) [11], is the definition of quality scenarios as part of the requirements-analysis workflow. Quality scenarios define measures that serve to validate the requirement to which they are associated. Furthermore, they specify the context in which the measurement process is to take place (*e.g.*, it is not the same measuring performance with 10 simultaneous users than with 10,000). Quality scenarios turn requirements into measurable requirements.

In order to model these measurable requirements, in this paper we extend $i^*$DWRA. The result of this extension is the $i^*$-based *data-warehousing quality measurement* framework ($i^*$DWQM). An important advantage of this framework is that, to our knowledge extent, it is the first proposal that traces quality scenarios back to the originating stakeholders' needs. Also, our proposal stresses the role of an often forgotten actor in data-warehouse development: the quality manager. Quality managers are responsible for orchestrating and leveraging the different stakeholders' interests during the data-warehouse development. Such interests include certain quality restrictions that must be respected for the project to be considered successful. It is important to note how, contrary to other measures proposed by different authors for $i^*$ diagrams [3, 4], our emphasis is not on the quality of the diagram *per se*, but on the provision of mechanisms to model the levels of quality required by the system under development.

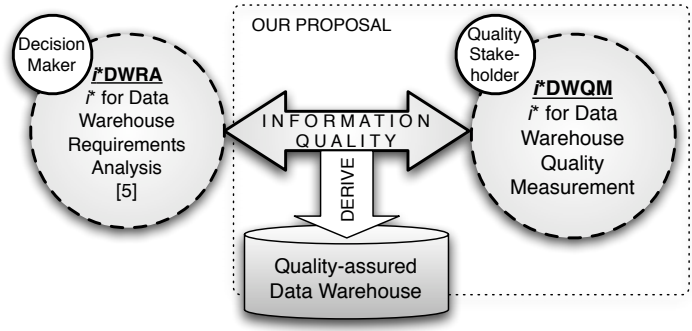

**Fig. 1.** Adding measurable quality scenarios for data-warehouse requirements analysis

The remainder of the paper is structured as follows: we present $i^*$DWQM next (§2) as an extension of $i^*$DWRA (see Fig. 1) that, for the sake of understandability, is also sketched. Both the measurement concepts and the chosen notation are then further illustrated with a sample application (§3). Our proposal has been defined by using the *unified modelling language* (UML) [14] profiling capabilities, which has permitted us to implement it in the *Eclipse*[1] development platform (§4). Finally, we summarise the main contributions of this paper and outlines some future lines of research (§5).

## 2 Modelling Framework for Data-warehouse Quality Measurement

Empirical research shows that the definition of measures and the description of measuring efforts in literature suffer from the typical symptoms of any relatively young discipline [1] and present many flaws that compromise their completeness and consistency. In order to overcome these problems, in [5] a *software measurement ontology* (SMO) has been proposed. Until the new ISO/IEC 25000 standard series appear[2], SMO reflects a compromise solution to solve the many inconsistencies and gaps detected in standards and research proposals. We have followed this ontology for the definition of $i^*$DWQM, in order to facilitate its adoption in the measurement domain. For the sake of understandability, in Appendix A the definitions of the ontology terms that have been used along this paper are formally reproduced. Interested readers may find further information about the whole ontology in [5].

### 2.1 Modelling Requirements Analysis with $i^*$DWRA

As it has been aforementioned, $i^*$DWRA is a data-warehouse requirements analysis framework that has proven useful for the discovery of data-warehouse requirements out of business goals. This purpose is achieved by identifying the decisions that decision makers usually are faced to. The $i^*$DWRA framework has been defined in two steps; first, in [12], a UML profile for $i^*$ (the $i^*$ profile, see Table 1, col. 4 & 5) has been provided. This profile elegantly redefines $i^*$ concepts and relationships [18] in terms of UML modelling elements. These elements permit to model both the organisational context (by means of the $i^*$ *strategic dependency* (SD) diagram) and the actors' rationale (by means of the $i^*$ *strategic rationale* (SR) diagram) when interacting with the data warehouse. Such concepts include *intentional elements* –actors (◯), goals (◯), tasks (◇), softgoals (◠), resources (▭), and beliefs (◌)– and *intentional relationships* –intentional dependencies (-▷-), means-end relationships (-▷), task-decompositions (—+), and contributions ( → ).

Over this $i^*$ profile, the second step has consisted in adding specific semantics for data-warehouse requirements analysis (see Table 1, col. 1–3). For the sake of simplicity, in this table only the $i^*$ elements that have been extended by the $i^*$DWRA framework are listed.

---

[1] URL: `www.eclipse.org`
[2] Namely, *software product quality requirements and evaluation* (SQuaRE)

Table 1. Mapping $i^*$DWRA concepts into the $i^*$ framework

| Analysis Concept | $i^*$DWRA | | $i^*$ Profile | |
|---:|:---:|:---|:---:|:---:|
| | Stereotype | Notation | Stereotype | UML |
| Strategic Goal | Strategy | ◯ + «strategy» | Goal | Class |
| Decisional Goal | Decision | ◯ + «decision» | Goal | Class |
| Informational Goal | Information | ◯ + «information» | Goal | Class |
| Info. Requirement | Requirement | ◯ + «task» | Task | Class |
| Context | Resource | ☐ + «context» | Resource | Class |
| Measure | Resource | ☐ + «measure» | Resource | Class |

Let us now give an example to illustrate how to properly read Table 1: Let us assume that we wish to model a data warehouse information requirement (see col. 1) in $i^*$DWRA. For this task, we would have to use the `Requirement` stereotype (col. 2). This stereotype provides additional semantics and notation (col. 3) to the $i^*$ task concept. $i^*$ tasks are mapped into UML by means of the `Task` stereotype (col. 4) on the UML `Class` modelling element (col. 5).

This modelling framework has been the basis on which we have performed a further extension to permit the definition of quality scenarios, which we have called $i^*$ *data-warehouse quality measurement* ($i^*$DWQM) framework.

It is worth noting that the `Measure` concept defined in $i^*$DWRA refers to the *analysis measures* employed during the decision-making process supported by the data-warehouse. Therefore, it must not be confused with the measure concept that appears in the context of data-warehouse quality scenarios, which we will explore next.

### 2.2  Modelling Quality Scenarios with $i^*$DWQM

Table 2. Mapping of SMO concepts into $i^*$DWQM

| SMO Concept | Equivalent $i^*$ Element |
|:---|:---:|
| Indicator, Derived Measure, Base Measure | Goal |
| Analysis Model, Measurement Function, Measurement Method | Task |
| Entity Class, Decision Criteria | Resource |
| Attribute | Belief |

The $i^*$DWQM framework enriches $i^*$DWRA with the capability of specifying quantifiable quality scenarios for the assurance of quality requirements associated with data-warehouses. As we have aforementioned, our proposal is based on SMO [5], in order to facilitate its understandability and help in its adoption by quality stakeholders.

*Mapping Quality Stakeholders into Actors.* In order to model quality stakeholders in $i^*$DWQM, following the $i^*$DWRA proposal, we use the $i^*$ actor modelling element. For instance, quality managers are actors that are in charge of defining quantifiable quality scenarios.

*Modelling Quality Scenarios for Data Warehouses.* $i^*$DWQM establishes a correspondence between particular SMO measurement concepts and more general $i^*$DWRA concepts. Such mapping is presented in Table 2. In this table, we can observe how SMO measures (`Indicator`s, `Derived Measure`s and `Base Measure`s) are mapped into measurement `Goal`s that can be achieved through certain `Task`s. These tasks are, namely, performing an `Analysis Model`, a `Measurement Function` or a `Measurement Method`, respectively. The measurement concepts `Entity Class` and `Decision Criteria` are mapped into `Resource`s, while the `Attribute` concept is mapped into a `Belief` in $i^*$DWRA.

Similarly, $i^*$DWQM maps the SMO relationships into $i^*$ intentional relationships in a hierarchical manner, as can be seen in Fig. 2.

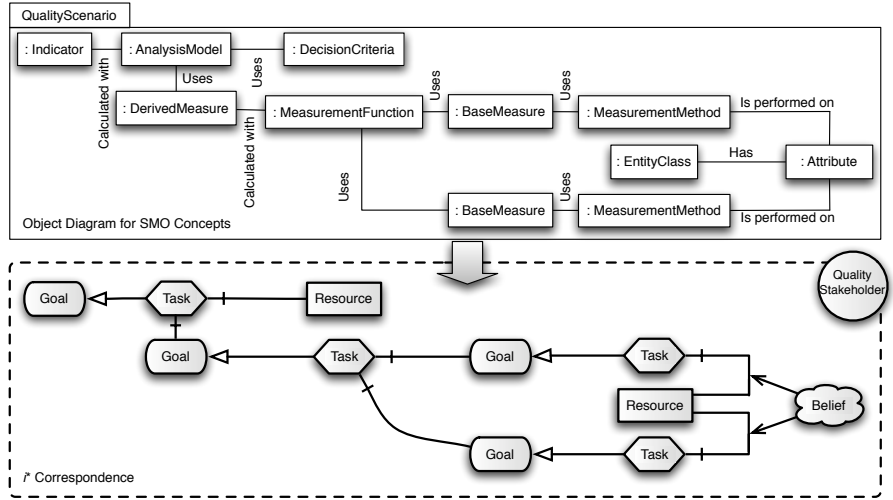

**Fig. 2.** Mapping a SMO occurrence for specifying quality scenarios with $i^*$DWQM

The upper part of this figure presents a SMO-based model that represents a generic quality scenario, while the lower part presents the equivalent $i^*$ model that has served to $i^*$DWQM as a basis for further enrichment. In this figure, we observe how, in SMO, an `Indicator` is related with an `Analysis Model`, which in turn has one or more `Decision Criteria` associated. An indicator is in fact a type of `Measure` that is made up of several other measures, be them `Derived Measure`s or `Base Measure`s. Derived measures are related with `Measurement Function`s, while base measures are associated with `Measurement Method`s. In SMO, analysis models, measurement functions and measurement methods can be related with `EntityClass`es through `Attribute`s.

The counterpart $i^*$DWQM relationships are presented in the lower part of Fig. 2. In this figure, we observe how indicators (mapped into goals) and analysis models (tasks) are related through a `Means-End` relationship. The same relationship appears between derived measures (goals) and their corresponding measurement functions (tasks) and base measures (goals) and their corresponding measurement methods (tasks). Another relevant relationship is that of `Task Decomposition` that appears between analysis models (tasks) and their related measures (goals) or decision criteria (resources), and also between measurement functions (tasks) and the associated measures (goals), or between measurement methods (tasks) and the associated entity classes (resources). Last, a `Contribution` relationship provides the attribute (belief) that permits the connection between measurement methods (tasks) and entity classes (resources). With this structure it is possible for quality managers to specify the quality scenarios associated with data-warehouse requirements.

*Modelling Dependencies among Stakeholders.* The quality scenarios modelled with $i^*$DWQM must be connected with particular non-functional requirements (softgoals) in the context of a particular data-warehouse informational scenario. Such information requirements and the associated softgoals provide respectively the context and the rationale for the measurement activity. As we have aforementioned, $i^*$DWRA information requirements are modelled as analysis `Tasks`. Analysis tasks may have different `softgoals` associated, which specify how the decision maker expects those tasks to be performed. At this point, existing quality models for data warehouses [15] are useful to choose among the set of non-functional requirements that are typical of this type of applications.

From the existing relationship between softgoals and measures, a dependency between the corresponding actors can be inferred. In $i^*$DWQM, this dependency is modelled as an intentional dependency from decision makers' *softgoals* (*depender*) to quality stakeholders' *goal* indicators (*dependee*) in order to achieve a given quality scenario *goal* (*dependum*) that a particular quality stakeholder knows how to measure. This modelling solution can be seen in Fig. 3.

*Modelling Measurement Attributes.* In the mapping presented so far, some SMO concepts, namely the measure characteristics `Unit Of Measurement`, `Scale`, and `Type Of Scale` are still missing. For the modelling of these concepts, $i^*$DWQM has made use of the *notes* mechanism enabled in $i^*$. Furthermore, the UML scaffolding that we have used to implement $i^*$DWQM (§4) further provides the necessary level of formalism to properly specify these SMO concepts.

## 3  Case Study

The case study chosen to illustrate our approach consists in a company selling automobiles across several countries. In this example (see Fig. 3), we have identified a *sales manager* as an actor that has several information requirements to be fulfilled by the data warehouse to be developed. During the requirements discovery phase, we have identified that *automobile sales be increased* is a strategic goal of the sales manager. From this strategic goal, several decision goals have been derived: *sales price be decreased*, *promotions be determined*, and so on. Focusing on the first decision goal, we have obtained two information goals: *automobile price be analysed* and *automobile sales be analysed*. Concerning the first one, we have recognised that the information requirement *analyse automobile sale price* is the means for achieving this decision goal, and for this analysis, the sales manager needs to check the prices and automobiles as fact and dimensions of the data warehouse analysis, respectively.

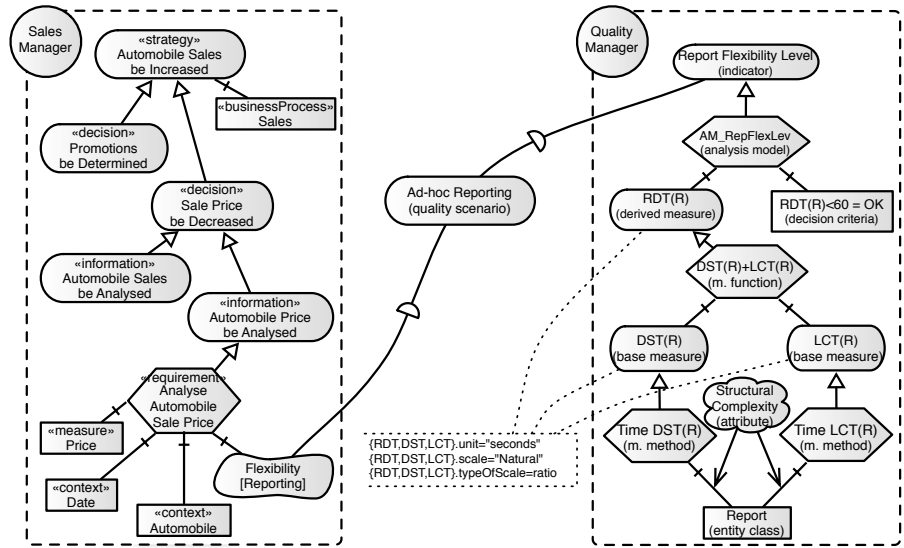

**Fig. 3.** $i^*$DWQM model for the *ad-hoc* reporting quality scenario

In addition, the analysis of the automobile sales price also needs the system to be flexible, where by flexible we refer to the extent to which the data-warehouse software facilitates *ad-hoc* reporting [15] (see Fig. 3). The quality scenario associated with this softgoal has been defined by the *quality manager* as follows: "A sales manager is able to design the required report, based on her mental model of the data warehouse, in less than 60 seconds" (referred to as "Ad-hoc Reporting" in Fig. 3). In Fig. 3, this quality scenario has been modelled with the aid of a *report flexibility level* indicator that evaluates the time it takes to the sales manager to design reports *ad hoc*. This indicator relies on a *derived measure* called *report design time* (`RDT(R)`) that, measured over a given report, returns the number of seconds that it takes to the sales manager to actually design the report. The quality scenario establishes that no more than 60 seconds is an acceptable time interval. This fact is captured in the *decision criteria* associated with the indicator. This measure is calculated through the sum of two base measures (measurement function), namely the *data selection time* (`DST(R)`) and the *layout composition time* (`LCT(R)`). These measures are assigned values by applying the corresponding measurement method, which consists in both cases in timing the corresponding tasks over a given report (the entity class). The belief in Fig. 3 indicates that these measures evaluate the structural complexity attribute associated with the report. Last, the unit of measurement, scale, and type of scale concepts are specified as additional notes in Fig. 3.

## 4   Implementation

The $i^*$DWQM has been implemented as an extension of UML and has been deployed in the *Eclipse* development platform (Fig. 4). Specifically, UML provides a standard extension formalism: UML profiles. These profiles consist of a set of stereotypes for particular UML modelling elements and some related tag definitions and constraints that, together, permit UML to host our modelling language. The $i^*$DWQM profile is based on two preexisting UML profiles for modelling $i^*$ diagrams adapted to the data-warehousing discipline, *i.e.*, the $i^*$DWRA and the $i^*$ profile [12] (see Fig. 4). So far, we have presented the mappings that support the definition of the necessary stereotypes for properly representing the $i^*$DWRA modelling elements in UML. While some concepts have been directly mapped to $i^*$ elements, others (namely, `Unit Of Measurement`, `Scale`, and `Type Of Scale` concepts), which in a pure $i^*$ framework can be modelled as notes, have been implemented in our profile as tag definitions associated with the measurement-related stereotypes. In addition to the modelling elements, $i^*$DWQM also considers the required constraints (derived from SMO) that assure the right use of these stereotypes, *e.g.*, forcing that only analysis models be the means for achieving an indicator. In this way, we provide a coherent modelling environment for (i) analysing data-warehouse requirements and (ii) associating a quantitative means (through measures) to assess their quality.

## 5   Conclusions

In this paper, we have presented $i^*$DWQM, a modelling framework to specify measurable quality scenarios that contribute to the assessment of the quality with which data-warehouse requirements are achieved. The completeness and unambiguity of the framework is facilitated by the use of a well-known Software Measurement Ontology [5] for its definition. Moreover, the UML scaffolding on which our approach is based contributes to achieving the desired degree of portability.

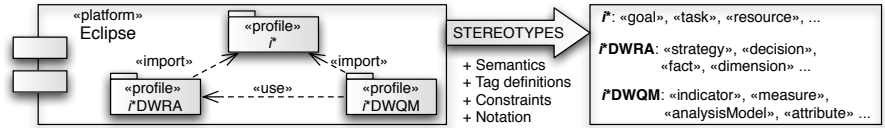

**Fig. 4.** The implemented UML profiling architecture for modelling with $i^*$DWQM

The use of our framework complements existing goal-oriented approaches for the development of data warehouses with several additional advantages:

- It increases the weight of quality scenarios and quality managers in the modelling process.
- It adds emphasis to the, often forgotten, measurable aspect that should be always associated with requirements in order to decrease the risks associated with system development.
- It provides a means to reason about how such measurement should take place, with the final goal of orchestrating and leveraging the different stakeholders' interests during the data-warehouse development.
- It provides traceability between the quality scenarios and the particular stakeholders' needs

Additionally, the measurement domain also obtains benefits out of being associated with goal-oriented approaches, among which we would like to stress out the provision by these approaches of a much richer organisational context than the one provided by the SMO for the definition of measures.

Although this framework has been devised for its application to data warehouses, their characteristics make us believe that $i^*$DWQM can be equally useful for other domains. This hypothesis constitutes one of our future lines of research. Last but not least, measuring models open the path for model-driven data-warehouse development frameworks (see *e.g.* [13]) to take them into account for the automatic generation of application tests.

## 6 Acknowledgements

This work has been supported by the projects: TIN208-00444, ESPIA (TIN2007-67078) from the Spanish Ministry of Education and Science (MEC), QUASIMODO (PAC08- 0157-0668) from the Castilla-La Mancha Ministry of Education and Science (Spain), and DEMETER (GVPRE/2008/063) from the Valencia Government (Spain). Jose-Norberto Mazón and Jesús Pardillo are funded by MEC under FPU grants AP2005-1360 and AP2006-00332, respectively.

Goal-oriented Data Warehouse Quality Measurement     11## A    Software Measurement Ontology Terms Definition

Table 3. Excerpt of SMO term definitions employed in this paper

| Concept | Definition |
| --- | --- |
| Entity Class | The collection of all entities that satisfy a given predicate |
| Attribute | A measurable physical or abstract property of an entity, that is shared by all the entities of an entity class |
| Scale | A set of values with defined properties |
| Type of Scale | The nature of the relationship between values on the scale |
| Unit of Measurement | Particular quantity, defined and adopted by convention, with which other quantities of the same kind are compared in order to express their magnitude relative to that quantity |
| Base Measure | A measure of an attribute that does not depend upon any other measure, and whose measurement approach is a measurement method |
| Derived Measure | A measure that is derived from other base or derived measures, using a measurement function as measurement approach |
| Indicator | A measure that is derived from other measures using an analysis model as measurement approach |
| Measurement Method | Logical sequence of operations, described generically, used in quantifying an attribute with respect to a specified scale. (A measurement method is the measurement approach that defines a base measure) |
| Measurement Function | An algorithm or calculation performed to combine two or more base or derived measures. (A measurement function is the measurement approach that defines a derived measure) |
| Analysis Model | Algorithm or calculation combining one or more measures with associated decision criteria. (An analysis model is the measurement approach that defines an indicator) |
| Decision Criteria | Thresholds, targets, or patterns used to determine the need for action or further investigation, or to describe the level of confidence in a given result |